\documentclass[reqno]{JHEP3}
\usepackage{graphicx}
\usepackage{amsmath}
\newcommand{\abs}[1]{\left\vert#1\right\vert}

\newcommand{\alphaprime}{\alpha '}
\newcommand{\zbar}{\overline{z}}
 
\DeclareMathOperator{\Tr}{Tr}
\title{Induced Gravity on Intersecting Branes}

\author{Friedel T.J. Epple\\Centre for Mathematical Sciences\\
Wilberforce Road\\Cambridge CB3 0WA, UK\\email:
\email{F.Epple@damtp.cam.ac.uk}}

\abstract{We establish Einstein-Hilbert gravity couplings in the
effective action for Intersecting Brane Worlds. The
four-dimensional induced Planck mass is determined by calculating
graviton scattering amplitudes at one-loop in the string
perturbation expansion. We derive a general formula linking the
induced Planck mass for $\mathcal{N}=1$ supersymmetric backgrounds
directly to the string partition function. We carry out the
computation explicitly for simple examples, obtaining analytic
expressions.}  \preprint{DAMTP-2004-82}

\begin{document}

\section{Introduction}

Intersecting Brane Worlds are among the most promising candidates
for constructing consistent string theory vacua with a realistic
low-energy limit. They are usually constructed from D6-branes
which intersect on a four-manifold and wrap different three-cycles
in the transverse compact space \cite{IBW, SusyOfolds} (for recent
reviews and more references see \cite{Lust:2004ks,
Kiritsis:2003mc}). Open strings stretching from one brane to
another give rise to chiral fermions which are localized on the
intersection locus due to the finite string tension. Interactions
of the open string fields have been studied in some detail, for
instance with respect to gauge coupling threshold corrections
\cite{Lust:2003ky, Blumenhagen:2003jy}, Yukawa couplings
\cite{Cvetic:2003ch, Cremades:2003qj, Abel:2003vv, Abel:2003yx}
and the tachyon potential \cite{Hashimoto:2003xz, Jones:2003ew,
Epple:2003xt, Nagaoka:2003zn}. On the other hand, gravitational
couplings have received much less attention.\footnote{Interaction
terms for the geometric moduli of toroidal type IIB orientifolds
have recently been computed in \cite{Lust:2004cx}} It is clear
that knowing their exact form will be crucial in determining the
cosmological evolution of Intersecting Brane Worlds and also for
studying low energy phenomenology. The simplest non-trivial
gravity coupling is an induced Einstein-Hilbert term on the
intersection locus. We demonstrate that such a term arises at
one-loop level in the string perturbation series and show how to
compute its magnitude. Although in this article we are only
concerned with the Einstein-Hilbert term, one-loop computations
like ours will have other important applications. For example, in
the context of moduli stabilization, it will be interesting to
determine one-loop corrections to the scalar potential in string
compactifications.

Our interest in induced Einstein gravity is partly motivated by
the observation by Dvali, Gabadadze and Porrati (DGP)
\cite{Dvali:2000hr, Dvali:2000xg} that an effective gravity action
of the form

\begin{equation}
S=M^{D-2}\int d^Dx \sqrt{-g} \mathcal{R}^{(D)}+M_P^2\int_{\Sigma}
d^4x \sqrt{-\tilde{g}} \mathcal{R}^{(4)}
\end{equation}

leads to localized four dimensional gravity on the subspace
$\Sigma$ of a flat $D$-dimensional bulk, irrespective of the
actual volume of the extra dimensions. The first term in the above
expression is the Einstein-Hilbert action for the bulk which is
constructed from the $D$-dimensional metric $g$ whereas the second
term is the Einstein-Hilbert on $\Sigma$ which is constructed from
the pullback metric $\tilde{g}$. For an infinitely thin
subspace\footnote{We deliberately avoid the term brane because in
the context of this paper, the subspace $\Sigma$ will be the
intersection locus of branes at angles} $\Sigma$, the localization
was shown to be perfect if $D\geq 6$. On the other hand, if one
assumes an effective width for the subspace in question, the bulk
space becomes infrared transparent leading to $D$-dimensional
gravity at ultra-low energies. This is in sharp contrast to the
Kaluza-Klein scenario where extra dimensions become visible at
high energies.

\FIGURE{
  \includegraphics{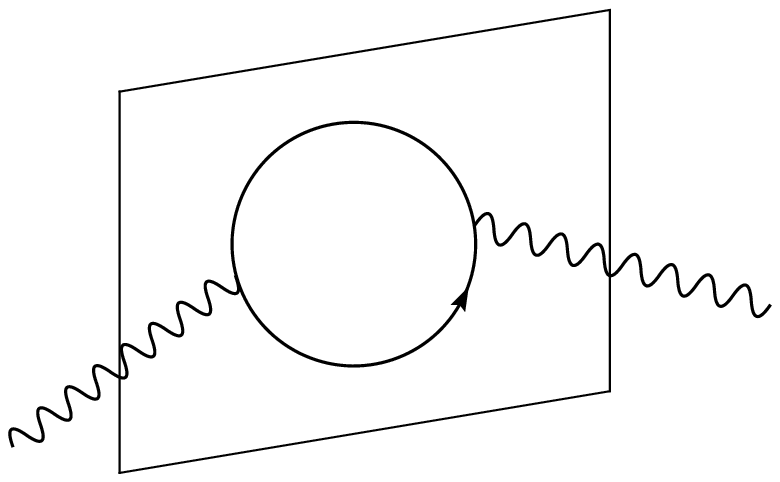}
  \caption{Localized matter loops on a surface $\Sigma$ renormalize the graviton two-point function.}
}

The assumptions leading to the DGP model are in fact extremely
generic. An Einstein-Hilbert term on a subspace $\Sigma$ can
easily be generated from interactions of localized matter with
bulk gravitons. From a field theory perspective, the graviton
propagator receives corrections from diagrams with matter on the
brane running in loops. In a purely four-dimensional spacetime,
these induce an Einstein-Hilbert term whenever the matter fields
are non-conformal \cite{Adler:1982ri}. Therefore, we can expect a
non-vanishing four-dimensional Planck mass in all but the most
special circumstances. In order to realize DGP's scenario in
string theory, one needs to find configurations where localized
matter leads to an induced Einstein-Hilbert action. The simplest
setup with localized matter is a single type II D-brane for which
the gravitational couplings have been computed at tree level in
reference \cite{Bachas:1999um}. In this particular case the large
supersymmetry prevents an Einstein-Hilbert term and the induced
gravity action starts with terms which are quadratic in curvature.
Similarly, a single orientifold plane - on which twisted closed
strings are localized - does not lead to DGP-gravity (see
\cite{Schnitzer:2002rt} for a tree-level computation of the
effective action). On the other hand, string theory realizations
of induced gravity have been found for a variety of backgrounds.
Orbifolds \cite{Kiritsis:1995ta, Kiritsis:2001bc} and, more
generally, Calabi-Yau compactifications \cite{Antoniadis:1997eg,
Antoniadis:2002tr, Antoniadis:2003sw} have been considered as well
as non-compact orientifolds of type IIB string theory
\cite{Kakushadze:1998tr, Kakushadze:2001bd, Kohlprath:2003pu}.
From a phenomenological point of view, the latter are interesting
because the requirement of a crystallographical action of the
orbifold group can be dropped and one has more freedom in
constructing GUT models. For the specific case of non-compact
$\mathbb{Z}_N$ orientifolds with $D3$-branes, the induced
four-dimensional Planck mass has been obtained explicitly
\cite{Kohlprath:2003pu} from a string theory computation.

The purpose of this paper is to establish induced gravity on
intersecting branes. Indeed, the low-energy effective action for
intersecting brane worlds generically has running gauge couplings
and is therefore non-conformal. Thus, from a four-dimensional
point of view, we can expect to find an induced Einstein-Hilbert
term on the intersection locus. In this paper, we only consider
intersecting brane worlds with at least $\mathcal{N}=1$
supersymmetry in four dimensions because these provide the only
known examples in which both RR- and NSNS-tadpoles are cancelled,
enabling us to perform one-loop computations. In section
\ref{graviton scattering}, we show how the induced Planck mass can
be computed in this class of models. In section \ref{examples}, we
explicitly carry out the computation for supersymmetric
$\mathbb{Z}_N$ orientifolds of $T^6$ with local tadpole
cancellation. The results we obtain are simple analytical
expressions. The induced four-dimensional Planck mass generically
receives contributions which are independent of the
compactification scale $R$. Curiously, we find that for even order
orbifold groups, the induced Planck mass additionally receives
corrections of the order $R^2$ if one of the internal tori is
taken to be large. These can be interpreted as coming from a
Kaluza-Klein reduction of a six-dimensional induced Planck mass.
In section \ref{conclusions}, we discuss our results.

\section{Graviton scattering}
\label{graviton scattering}

We obtain the effective gravity action using the S-matrix
approach, i.e. by computing the amplitude for a graviton
scattering off the brane intersection locus and comparing the
result to a field theory ansatz. Since we are interested in the
effective field theory on the worldvolume of the brane
intersection we have to use string worldsheets which are localized
correspondingly. The lowest order worldsheet which satisfies this
requirement is the annulus which has its two boundaries attached
to two different branes. We obtain the scattering amplitude by
inserting two graviton vertices in the path integral. However, the
annulus alone gives a divergent expression for the scattering
amplitude. This is because the D-branes carry tadpoles which lead
to divergences when the corresponding fields propagate in the
closed string channel. At one-loop level, it is therefore
impossible to consider two intersecting branes independently from
the background they are in. Instead, we always have to work with
the full set of one-loop surfaces. For string vacua with compact
extra dimensions, uncancelled RR-tadpoles imply inconsistent
equations of motion for the RR p-forms \cite{Polchinski:1996na}.
Thus, their cancellation is a necessary consistency check for any
such open string vacuum. On the other hand, NS-NS tadpoles merely
hint at a perturbative instability of the chosen vacuum. Thus, for
some time they had been largely ignored in D-brane model building.
However, any perturbative loop computation is rendered meaningless
if there are uncancelled tadpoles, of any kind, and when computing
loop corrections to scattering amplitudes, we are forced to
consider only perturbatively consistent and stable backgrounds.

The only intersecting brane worlds with complete tadpole
cancellation we know of are supersymmetric orientifolds
\cite{SusyOfolds, Blumenhagen:1999md, Blumenhagen:1999ev,
Blumenhagen:2004di}. Just like for the vacuum amplitude,
divergencies in the graviton scattering amplitude should be
cancelled between the annulus, Klein bottle and M\"obius strip.
Also, from the orbifold section we have twisted closed strings
which contribute to the induced Einstein-Hilbert term through the
torus worldsheet. The details of the string computation are
similar to those presented in \cite{Kohlprath:2003pu} where
induced gravity in standard orientifolds with $D3$-branes was
investigated, i.e. in orientifolds with projection $\Omega J$
where $J$ acts on the complexified coordinates of the extra
dimensions as $J:Z_i\mapsto -Z_i$. Contrastingly, orientifolds
with intersecting branes arise from an orientifold projection
$\Omega\mathcal{R}$ where $\mathcal{R}$ acts as
$\mathcal{R}:Z_i\mapsto \overline{Z}_i$. The torus amplitude is
unaffected by the orientifold projection. Hence, the result that
the torus contribution to the four-dimensional Einstein-Hilbert
term is proportional to the Euler number of the compactification
space \cite{Antoniadis:1997eg} carries over to orientifold models
with intersecting branes. The difference between orientifolds with
intersecting branes and the models presented in
\cite{Kohlprath:2003pu} is what might be called the orientifold
sector, i.e. the contributions from the Klein bottle, annulus and
M\"obius strip. We find that this difference leads to considerable
complications. In particular, it is not always obvious how (if at
all) tadpole cancellation guarantees the absence of UV-divergences
in the graviton scattering amplitude. However, for the simplest
case in which tadpoles are cancelled locally by placing an
appropriate number of D-branes on top of the orientifold planes,
we can carry out the computation explicitly and obtain a finite
answer. First, we will outline the general procedure for computing
the relevant scattering amplitudes with special emphasis on
correctly normalizing the path integral and apply our results to
concrete models in section \ref{examples}.

\subsection{Vertex operators, propagators and path integral
normalization}

We are interested in contributions to the graviton scattering
amplitude coming from the annulus ($\mathcal{A}$), Klein bottle
($\mathcal{K}$) and M\"obius strip ($\mathcal{M}$). In the path
integral approach, they are given by:

\begin{equation}
\label{string amplitude} A_{\sigma}=\int d^2z_1 \int d^2z_2
\langle
V_{p_1,\varepsilon_1}(z_1,\zbar_1)V_{p_2,\varepsilon_2}(z_2,\zbar_2)\rangle_{\sigma};\qquad
\sigma = \mathcal{A}, \mathcal{K}, \mathcal{M}
\end{equation}

where $\langle\dots\rangle_{\sigma}$ denotes the Polyakov path
integral for the NSR string on a Riemann surface $\sigma$. The
path integral includes summation over different spin structures
$s$ and metric moduli $t$. As we are dealing with a CP-even
scattering process, only the even spin structures contribute to
the amplitude and we can therefore use the graviton vertex in the
$(0,0)$-picture:

\begin{equation}
\label{vertex}
V_{p,\varepsilon}(z,\overline{z})=g_c\varepsilon_{\mu\nu}
:(\partial X^{\mu}+i\psi^{\mu}p\cdot\psi)(z)
(\overline{\partial}X^{\nu}-i\tilde{\psi}^{\nu}p\cdot\tilde{\psi})(\zbar)e^{ip\cdot
X(z,\zbar)} :.
\end{equation}

Here, $g_c$ is the closed string coupling and we have set
$\alphaprime=2$. Since we are interested in the overall magnitude
of the graviton scattering amplitude, the correct normalization of
the path integral is crucial. For the surfaces in question, it is
unambiguously given by

\begin{equation}
\langle
1\rangle_{\text{one-loop}}=\int_0^{\infty}\frac{dt}{2t}\Tr\left(P_{\text{orientifold}}
P_{GSO}e^{-2\pi t\alphaprime H}\right),
\end{equation}

as can be checked by applying the Coleman-Weinberg formula
\cite{Coleman:1973jx} to the string theory particle content.
Expanding the orientifold projection $P_{\text{orientifold}}$, one
obtains the individual contributions from all four one-loop
surfaces. The normalization of the vertex operator \eqref{vertex}
is the same for all topologies and spin structures, it defines the
string coupling \cite{Polchinski:1998rq}.

All three surfaces $\mathcal{A},\mathcal{K}$ and $\mathcal{M}$ in
the orientifold sector can be obtained from a covering torus by an
identification under an appropriate anti-holomorphic involution

\begin{equation}
I_{\mathcal{A}}, I_{\mathcal{K}}: z\rightarrow 1-\zbar,\qquad
I_{\mathcal{M}}: z\rightarrow 1-\zbar+\frac{\tau}{2}.
\end{equation}

The complex modulus $\tau$ of the covering torus is related to the
loop modulus $t$ via

\begin{equation}
\label{moduli}
\tau=
  \begin{cases}
    it/2 & \mathcal{A}\\
    2it & \mathcal{K}\\
    1/2+it/2 & \mathcal{M}
  \end{cases}
\end{equation}

The propagators of the worldsheet fields $X$ and $\psi$ on a
surface $\sigma$ with a particular spin structure $s$ can be
derived from the corresponding expressions for the covering torus
by using the method of images \cite{Burgess:1987wt}. We write them
as operator equations:

\begin{align}
\partial X^{\mu}(z)\partial X^{\nu}(w)
&=\eta^{\mu\nu}(\partial_z\partial_wP_B(z,w;\tau)+\pi/4\tau_2))\\
\partial X^{\mu}(z)\overline{\partial}
X^{\nu}(\overline{w})
&= \eta^{\mu\nu}(\partial_z\partial_{\overline{w}}P_B(z,I_{\sigma}(w);\tau)-\pi/4\tau_2)\\
\psi^{\mu}(z)\psi^{\nu}(w)
&= \eta^{\mu\nu} P_F(z,w;\tau,s)\\
\psi^{\mu}(z)\tilde{\psi}^{\nu}(\overline{w})
&= \eta^{\mu\nu} P_F(z,I_{\sigma}(w);\tau,s)\\
\tilde{\psi}^{\mu}(\zbar)\tilde{\psi}^{\nu}(\overline{w}) &=
\eta^{\mu\nu}
\overline{P}_F(\zbar,\overline{w};\tau,\overline{s}),
\end{align}

where $\sigma=\mathcal{A}$, $\mathcal{K}$, $\mathcal{M}$ denotes
the surface and $P_B$, $P_F$ are the bosonic and fermionic
propagators on the torus

\begin{gather}
P_B(z,w;\tau)=-\frac{1}{4}\ln\abs{\frac{\vartheta_1(z-w|\tau)}{\vartheta_1'(0|\tau)}}^{2}+\frac{\pi(z_2-w_2)^2}{2\tau_2}\\
P_F(z,w;\tau,s)=\frac{i}{2}\frac{\vartheta_s(z-w|\tau)}{\vartheta_1(z-w|\tau)}\frac{\vartheta_1'(0|\tau)}{\vartheta_s(0|\tau)},
\end{gather}

involving Jacobi's theta functions (for definitions and selected
properties, see appendix \ref{Jacobi}). By 'operator equation' we
mean that an equation holds inside a path integral. For instance,
we have
$\langle\psi^{\mu}(z)\psi^{\nu}(w)\rangle=\langle\eta^{\mu\nu}
P_F(z,w;\tau,s)\rangle$. Note that we cannot simply drop the
brackets on the right hand side because the Polyakov path integral
is not normalized to unity, $\langle 1\rangle\neq1$. There is
another subtlety regarding the overall normalization of the
scattering amplitude having to do with the integration of bosonic
zero-modes. These appear exponentiated in the vertex operators.
Since normal ordering only affects the oscillator modes, the
graviton scattering amplitude contains a factor of $\langle
e^{i(p_1+p_2)x_0}\rangle$ in addition to all full contractions
between the oscillator parts of the worldsheet fields. Here,
$x_0^{\mu}$ is the zero-mode of the worldsheet scalars. If we
denote by $\langle\dots\rangle_*$ the path integral with
contributions from $x_0^{\mu}$ excluded ($\mu=0\dots 3$), we find:

\begin{equation}
\langle e^{i(p_1+p_2)x_0}\rangle=\int d^4x_0
e^{i(p_1+p_2)x_0}\langle 1\rangle_*
=(2\pi)^4\delta^{(4)}(p_1+p_2)\langle 1\rangle_*
=(2\pi)^4\delta^{(4)}(p_1+p_2)\frac{1}{V_4}\langle 1\rangle,
\end{equation}

where $V_4$ is the regulated four-dimensional volume. Using this
identity, we can directly relate the two graviton scattering
amplitude to the vacuum amplitude. Before doing so, we introduce
one more piece of notation:

\begin{equation}
\langle 1\rangle_{\sigma}
=\sum_s\int_0^{\infty}\frac{dt}{2t}Z_{\sigma}(t,s).
\end{equation}

Also, for notational convenience, we will let a product of two
fields stand for its contraction (propagator) whenever it appears
outside the path integral. Our discussion of the path integral
normalization also applies to the work in \cite{Kohlprath:2003pu}
where the normalization of the scattering amplitudes was left
undetermined.

\subsection{One-loop renormalization of the four-dimensional Planck mass}
\label{renormalization}

Now we have all the ingredients for the computation of the
graviton scattering amplitude \eqref{string amplitude}. Because we
are interested in supersymmetric backgrounds, the one-loop vacuum
amplitude vanishes when summed over spin structures. Therefore, we
can only get a non-vanishing contribution to the scattering
amplitude if we contract at least one pair of fermionic fields.
Looking at the graviton vertex, we see that this already amounts
to the correct power of momenta for the induced Einstein-Hilbert
term\footnote{We also see that, as expected, the four-dimensional
cosmological constant does not receive corrections at one-loop and
we are indeed dealing with a DGP type of gravity.} and therefore
we find that, taking into account the physical state conditions
for graviton vertices, the relevant contribution to the scattering
amplitude is given by:

\begin{align}
A=&(2\pi)^4\delta^{(4)}(p_1+p_2)\frac{1}{V_4}(-g_c^2)
\sum_{\sigma=\mathcal{A},\mathcal{K},\mathcal{M}}\sum_{s=2,3,4}
\int_0^{\infty}\frac{dt}{2t}Z_{\sigma}(t,s)\int d^2z_1\int d^2z_2 \nonumber\\
 & \times\frac{1}{4}\left\{\partial X\partial
X(\tilde{\psi}\tilde{\psi})_s^2-\overline{\partial}X\partial
X(\psi\tilde{\psi})^2_s
+\overline{\partial}X\overline{\partial}X(\psi\psi)^2_s -
\partial X\overline{\partial}X(\tilde{\psi}\psi)^2_s\right\}(p_1\varepsilon_2\varepsilon_1 p_2) ,
\end{align}

We should mention that this expression actually vanishes because
of four-dimensional momentum conservation and transversality of
the graviton polarization. For a completely strict computation, we
would have to either turn on momentum components in the transverse
directions (for which momentum conservation doesn't hold) or use
the three graviton scattering amplitude. However, both options
lead to considerably more complicated expressions. Instead of
following these prescriptions, we simply use the structure of the
kinetic term $(p_1\varepsilon_2\varepsilon_1p_2)$ to identify the
corresponding term in the effective field theory. In fact, it is
the only possible combination at quadratic order in the momenta
and is reproduced by the quadratic Einstein-Hilbert action on
Minkowski four-space:

\begin{equation}
S_{EH}=\frac{M_P^2}{2}\int d^4x
\left\{-\frac{1}{2}h_{\alpha\beta,\nu}h^{\nu\beta,\alpha}\right\}.
\end{equation}

Here, we have dropped terms in the action which vanish in
transverse traceless gauge because these do not contribute to the
tree level scattering amplitude. It is common practice to
parameterize the one-loop renormalization of the Einstein-Hilbert
term by a number $\delta$ which is related to the induced Planck
mass by $\delta=M_{P,\text{ind}}^{2}/2$. Thus, comparing the
coefficients of the kinetic term, we find that the one-loop
renormalization of the Planck-mass is given by

\begin{align}
\label{delta} \delta =& \frac{g_s^2}{V_4}
\sum_{\sigma=\mathcal{A},\mathcal{K},\mathcal{M}}\sum_{s=2,3,4}
\int_0^{\infty}\frac{dt}{2t}Z_{\sigma}(t,s)\int d^2z_1\int d^2z_2 \nonumber\\
 &  \times\frac{1}{4}\left\{\partial X\partial
X(\tilde{\psi}\tilde{\psi})_s^2-\overline{\partial}X\partial
X(\psi\tilde{\psi})^2_s
+\overline{\partial}X\overline{\partial}X(\psi\psi)^2_s -
\partial X\overline{\partial}X(\tilde{\psi}\psi)^2_s\right\},
\end{align}

which can be further simplified \cite{Kohlprath:2003pu} if one
uses the following identity for the fermionic propagators:

\begin{equation}
P_F(z,0;\tau,s)^2=-\partial_z^2\ln\vartheta_1(z|\tau)+\partial_v^2\frac{\vartheta_s(v|\tau)}{\vartheta_s(0|\tau)}|_{v=0}.
\end{equation}

The first term does not depend on the spin structure and can
therefore be dropped when dealing with supersymmetric backgrounds.
On the other hand, the second term does not depend on $z$ and
therefore can be pulled in front of the worldsheet integration in
\eqref{delta}. The remaining contribution from the bosonic
propagators was shown to be \cite{Antoniadis:1997vw}

\begin{equation}
\int d^2z_1\int d^2z_2 \left\{\partial X\partial
X(\tilde{\psi}\tilde{\psi})^2_s-\overline{\partial}X\partial
X(\psi\tilde{\psi})^2_s
+\overline{\partial}X\overline{\partial}X(\psi\psi)^2_s -\partial
X\overline{\partial}X(\tilde{\psi}\psi)^2_s\right\} =
\frac{\pi\tau_2}{4},
\end{equation}

where $\tau_2$ is the imaginary part of the complex modulus of the
covering torus. Altogether, using \eqref{moduli} the contribution
to the induced Planck mass from the orientifold sector becomes
$\delta=\delta_{\mathcal{A}}+\delta_{\mathcal{K}}+\delta_{\mathcal{M}}$
where the individual terms are given by

\begin{align}
\delta_{\mathcal{A}} &= \frac{g_s^2}{V_4} \partial_v^2
\sum_{s=2,3,4}\int_0^{\infty}
\frac{dt}{2t}Z_{\mathcal{A}}(t,s)\frac{\vartheta_s(v|it/2)}{\vartheta_s(0|it/2)}
\times \frac{\pi t}{8}\quad \Big|_{v=0}\\
\delta_{\mathcal{K}} &= \frac{g_s^2}{V_4}\partial_v^2
\sum_{s=2,3,4}\int_0^{\infty}
\frac{dt}{2t}Z_{\mathcal{K}}(t,s)\frac{\vartheta_s(v|2it)}{\vartheta_s(0|2it)} \times \frac{\pi t}{2}\quad \Big|_{v=0}\\
\delta_{\mathcal{M}} &= \frac{g_s^2}{V_4}\partial_v^2
\sum_{s=2,3,4}\int_0^{\infty}
\frac{dt}{2t}Z_{\mathcal{M}}(t,s)\frac{\vartheta_s(v|1/2+it/2)}{\vartheta_s(0|1/2+it/2)}\times
\frac{\pi t}{8}\quad \Big|_{v=0}.
\end{align}

In order to make potential UV-divergencies more visible, we
convert these expressions into tree channel using the following
modular transformations

\begin{align}
\mathcal{A}:&\quad \tau=it/2\rightarrow -\frac{1}{\tau}=il\\
\mathcal{K}:&\quad \tau=2it\rightarrow -\frac{1}{\tau}=il\\
\mathcal{M}:&\quad \tau=\frac{1}{2}+i\frac{t}{2}\rightarrow
-\frac{1}{\tau}\rightarrow-\frac{1}{\tau}+2\rightarrow
\left(\frac{1}{\tau}-2\right)^{-1}=-\frac{1}{2}+\frac{i}{2t}=il-\frac{1}{2}.
\end{align}

Then, using the modular transformation properties \eqref{modular1}
to \eqref{modular2} of Jacobi's theta functions, the individual
contributions become

\begin{align}
\delta_{\mathcal{A}}&=\frac{g_s^2}{V_4}
\partial_v^2\sum_{s=2,3,4} \int_0^{\infty} d\ell
\tilde{Z}_{\mathcal{A}}(\ell,s)\frac{\vartheta_s(-iv\ell|i\ell)}{\vartheta_s(0|i\ell)} \frac{\pi}{4\ell} \quad \Big|_{v=0}\\
\delta_{\mathcal{K}}&=\frac{g_s^2}{V_4}
\partial_v^2\sum_{s=2,3,4}\int_0^{\infty} d\ell
\tilde{Z}_{\mathcal{K}}(\ell,s)\frac{\vartheta_s(-iv\ell|i\ell)}{\vartheta_s(0|i\ell)} \frac{\pi}{4\ell} \quad \Big|_{v=0}\\
\delta_{\mathcal{M}}&=\frac{g_s^2}{V_4}
\partial_v^2\sum_{s=2,3,4}\int_0^{\infty} d\ell
\tilde{Z}_{\mathcal{M}}(\ell,s)\frac{\vartheta_s(-2iv\ell|-1/2+i\ell)}{\vartheta_s(0|-1/2+i\ell)}
\frac{\pi}{16\ell} \quad \Big|_{v=0},
\end{align}

where $\sum_s\int d\ell \tilde{Z}(\ell,s)$ is the vacuum amplitude
after transformation to tree channel. Here, $s$ denotes the spin
structure in the tree channel, whose relation to the loop channel
spin structure depends on the surface in question. Finally, this
simplifies to the following master equations:

\begin{align}
\label{treedelta1} \delta_{\mathcal{A}}&=-\frac{\pi
g_s^2}{4V_4}\partial_v^2\sum_{s=2,3,4}\int_0^{\infty} d\ell \ell
\tilde{Z}_{\mathcal{A}}(\ell,s)
\frac{\vartheta_s(v|i\ell)}{\vartheta_s(0|i\ell)} \quad\Big|_{v=0}\\
\delta_{\mathcal{K}}&=-\frac{\pi
g_s^2}{4V_4}\partial_v^2\sum_{s=2,3,4}\int_0^{\infty} d\ell \ell
\tilde{Z}_{\mathcal{K}}(\ell,s)
\frac{\vartheta_s(v|i\ell)}{\vartheta_s(0|i\ell)} \quad\Big|_{v=0}\\
\label{treedelta3} \delta_{\mathcal{M}}&=-\frac{\pi
g_s^2}{4V_4}\partial_v^2\sum_{s=2,3,4}\int_0^{\infty} d\ell \ell
\tilde{Z}_{\mathcal{M}}(\ell,s)
\frac{\vartheta_s(v|-1/2+i\ell)}{\vartheta_s(0|-1/2+i\ell)}
\quad\Big|_{v=0}.
\end{align}

All three contributions have the same numerical prefactor. This
raises the hope that tadpole cancellation would imply a UV-finite
one-loop correction of the Plank mass as was indeed the case for
the standard orientifolds in \cite{Kohlprath:2003pu}. However, it
turns out that UV-finiteness is much less obvious when it comes to
orientifolds with intersecting branes.


\section{Orientifolds with local tadpole cancellation}
\label{examples}

In supersymmetric four-dimensional orientifolds with intersecting
branes \cite{SusyOfolds, Blumenhagen:1999ev, Blumenhagen:2004di},
the transverse space is compactified on a Calabi-Yau 3-fold, on
which an appropriate orientifold projection acts. In the limit
where the Calabi-Yau becomes a toroidal orbifold, we can perform
explicit string computations. The orientifold projection combines
world-sheet parity with an antiholomorphic involution of the
orbifold space. The corresponding Klein bottle tadpoles can be
cancelled if one adds intersecting D6-branes. This cancellation
can be achieved either locally or globally. Configurations with
global tadpole cancellation (i.e. featuring branes which are not
fully aligned to the orientifold planes) have been particularly
successful in constructing phenomenologically appealing models.
Here, we will focus on models with local tadpole cancellation
which place intersecting branes on top of the orientifold planes
\cite{Blumenhagen:1999md, Blumenhagen:1999ev, Blumenhagen:2004di}.
The geometrical setup is the following: We consider a factorizable
six-torus $T^6=T^2\times T^2\times T^2$ and an orbifold group
$\mathbb{Z}_N=\{\Theta^n|n=1,\dots,N\}$ where $\Theta$ acts on the
torus coordinates as

\begin{equation}
\Theta:z_i\mapsto e^{2\pi i v_i}z_i, \quad i=1,2,3.
\end{equation}

The condition for preserved supersymmetry can be written $\sum_i
v_i=0 \mod 2\pi$. The orientifold projection is given by
$P=\tfrac{1}{2}(1+\Omega\mathcal{R})$ where $\mathcal{R}$ acts as
complex conjugation on all three two-tori. The shape of the tori
is essentially fixed by the requirement that the orbifold group
should act crystallographically. However, there are two possible
alignments of each torus with respect to the $\Omega\mathcal{R}$
orientifold planes. These are known as \textbf{A}- and \textbf{B}-
tori respectively and play an important role in ensuring
worldsheet consistency at one-loop level \cite{Blumenhagen:1999md,
Blumenhagen:2004di}.


\subsection{The $\mathbb{Z}_3$ orientifold}

The simplest non-trivial example of $\mathbb{Z}_N$-orientifolds is
the $\mathbb{Z}_3$-model which was first described in
\cite{Blumenhagen:1999ev, Pradisi:1999ii}. It has a twist vector
$v=(1,1,-2)/3$ and $M$ branes on top of each orientifold plane.
Orientifolds of type $\Omega\mathcal{R}$ don't have twisted
tadpoles and as a consequence, the orbifold acts on the Chan-Paton
indices through traceless matrices,

\begin{equation}
\Tr \gamma_k=0, \quad k=1\dots N-1.
\end{equation}


All possible choices for the three tori are allowed. For
simplicity, we will choose the \textbf{AAA}-lattice. The only
difference for a $\textbf{A}^i\textbf{B}^{3-i}$ torus would be an
overall factor of $3^i$ in the partition function coming from
additional brane intersections and $\mathcal{R}$-invariant
orbifold fixed points as well as from different momentum lattices
\cite{Angelantonj:1999xf, Blumenhagen:1999ev}. The various
contributions from the annulus, Klein bottle and M\"obius strip to
the partition function (in the tree channel) are given by:

\begin{align}
\mathcal{Z}_{\mathcal{A}}&=
\frac{M^2}{12\sqrt{3}}\frac{V_4}{(16\pi^2)^2}
\sum_{s=2,3,4}(-)^{s-1}\int_0^{\infty}d\ell
\frac{\vartheta_s(0|i\ell)}{\eta(i\ell)^{3}}
\Bigg\{\frac{\vartheta_s(0|i\ell)^3}{\eta(i\ell)^{9}}\mathcal{L}_{\mathcal{A}}(\ell)\nonumber\\
&\qquad\qquad\qquad\qquad
+3\sqrt{3}\prod_{i=1}^3\frac{\vartheta_s(v_i|i\ell)}{\vartheta_1(v_i|i\ell)}
-3\sqrt{3}\prod_{i=1}^3\frac{\vartheta_s(2v_i|i\ell)}{\vartheta_1(2v_i|i\ell)}
\Bigg\}\\
\mathcal{Z}_{\mathcal{K}}&=
\frac{16}{12\sqrt{3}}\frac{V_4}{(16\pi^2)^2}
\sum_{s=2,3,4}(-)^{s-1}\int_0^{\infty}d\ell\frac{\vartheta_s(0|i\ell)}{\eta(i\ell)^{3}}
\Bigg\{\frac{\vartheta_s(0|i\ell)^3}{\eta(i\ell)^{9}}\mathcal{L}_{\mathcal{K}}(\ell)\nonumber\\
&\qquad\qquad\qquad\qquad
+3\sqrt{3}\prod_{i=1}^3\frac{\vartheta_s(v_i|i\ell)}{\vartheta_1(v_i|i\ell)}
-3\sqrt{3}\prod_{i=1}^3\frac{\vartheta_s(2v_i|i\ell)}{\vartheta_1(2v_i|i\ell)}
\Bigg\}\\
\mathcal{Z}_{\mathcal{M}}&=
-\frac{8M}{12\sqrt{3}}\frac{V_4}{(16\pi^2)^2}
\sum_{s=2,3,4}(-)^{s-1}\int_0^{\infty}d\ell\frac{\vartheta_s(0|-1/2+i\ell)}{\eta(i\ell)^{3}}
\Bigg\{\frac{\vartheta_s(0|-1/2+i\ell)^3}{\eta(-1/2+i\ell)^{9}}\mathcal{L}_{\mathcal{M}}(\ell)\nonumber\\
&\qquad\qquad
+3\sqrt{3}\prod_{i=1}^3\frac{\vartheta_s(v_i|-1/2+i\ell)}{\vartheta_1(v_i|-1/2+i\ell)}
-3\sqrt{3}\prod_{i=1}^3\frac{\vartheta_s(2v_i|-1/2+i\ell)}{\vartheta_1(2v_i|-1/2+i\ell)}
\Bigg\}.
\end{align}

Here, $\mathcal{L}_{\sigma}(\ell)$ denotes lattice contributions
(from summing over Kaluza-Klein and winding modes) which rapidly
tend to unity as $l$ goes to infinity and whose exact form we
won't need. Tadpole cancellation therefore requires $M=4$. Knowing
the different contributions to the partition function, we can now
compute the induced Planck mass. Using the tadpole cancellation
condition, $M=4$, as well as equations \eqref{treedelta1} to
\eqref{treedelta3} and a variation of the Jacobi identity
(\ref{JacobiIdentity}, \ref{Jacobi1}), we find:

\begin{align}
\delta_{\mathcal{A}}&=+\frac{1}{2^7\pi^2}\int_0^{\infty}d\ell\ell\sum_{i=1}^3
\frac{\vartheta_1'(v_i|i\ell)}{\vartheta_1(v_i|i\ell)}\\
\delta_{\mathcal{K}}&=+\frac{1}{2^7\pi^2}\int_0^{\infty}d\ell\ell\sum_{i=1}^3
\frac{\vartheta_1'(v_i|i\ell)}{\vartheta_1(v_i|i\ell)}\\
\delta_{\mathcal{M}}&=-\frac{2}{2^7\pi^2}\int_0^{\infty}d\ell\ell\sum_{i=1}^3
\frac{\vartheta_1'(v_i|-1/2+i\ell)}{\vartheta_1(v_i|-1/2+i\ell)}.
\end{align}

Note that the four-dimensional volume $V_4$ drops out as required.
Adding up these contributions gives the simple expression

\begin{equation}
\delta=\frac{1}{2^6\pi^2}\int_0^{\infty}d\ell\ell\sum_{i=1}^3\left(
\frac{\vartheta_1'(v_i|i\ell)}{\vartheta_1(v_i|i\ell)}-\frac{\vartheta_1'(v_i|-1/2+i\ell)}{\vartheta_1(v_i|-1/2+i\ell)}\right).
\end{equation}

One immediately sees that the integral is UV-finite because in the
limit $l\rightarrow\infty$ the two different expressions in the
bracket rapidly approach the same constant. However, one can even
compute the integral explicitly, using special properties of the
Jacobi functions. The computation is carried out in appendix
\ref{usefulintegrals}, leading to the final result (reinserting a
factor of $1/\alphaprime=M_s^2$):

\begin{equation}
\delta=\frac{1}{2^{12}\pi^2\sqrt{3}}\left(\zeta(2,5/6)-\zeta(2,1/6)\right)g_s^2M_s^2\approx
5.0\times 10^{-4} g_s^2M_s^2.
\end{equation}

Here, $\zeta$ is the generalized or Hurwitz zeta function. Thus,
we get a finite, non-vanishing one-loop contribution to the
induced four-dimensional Planck mass. As could have been expected
from the localization of the relevant interactions, the induced
Planck mass is independent of the compactification scale. It is
also roughly of the order of the string mass. The remaining
orientifolds which were described in \cite{Blumenhagen:1999ev},
namely the $\mathbb{Z}_{4}$-, $\mathbb{Z}_{6}$- and
$\mathbb{Z}_{6}'$-models can be treated in the same manner. There
is, however, one modification coming from the fact that these
models contain a $\mathbb{Z}_{2}$ subgroup.

\subsection{Even order $\mathbb{Z}_N$-orientifolds}

If $\omega$ is the generator of a $\mathbb{Z}_2$ subgroup, then
its twist vector necessarily has a zero in one component, i.e. one
torus is invariant under the action of $\omega$. As a consequence,
the corresponding contribution to the vacuum loop amplitude
contains a factor of $\vartheta_1^2/\eta^6$ and a non-trivial
lattice factor from Kaluza-Klein and winding modes. As an
illustration of this point, we write down the computation of the
induced Planck mass for the $\mathbb{Z}_4$-case with type
\textbf{ABA} torus which has a twist vector $v=(1/4,1/4,-1/2)$.
The $\mathcal{A}$, $\mathcal{K}$ and $\mathcal{M}$ vacuum
amplitudes were calculated in \cite{Blumenhagen:1999ev}. In tree
channel, they are given by:

\begin{align}
\mathcal{Z}_{\mathcal{A}}&= \frac{M^2}{16}\frac{V_4}{(16\pi^2)^2}
\sum_{s=1}^4(-)^{s-1}\int_0^{\infty}d\ell\frac{\vartheta_s(0|i\ell)}{\eta(i\ell)^{3}}
\Bigg\{\frac{\vartheta_s(0|i\ell)^3}{\eta(i\ell)^{9}}\mathcal{L}_{\mathcal{A}}(\ell)\nonumber\\
&\qquad\qquad - \frac{\vartheta_s(0|i\ell)}{\eta(i\ell)^3}
\frac{\vartheta_s(1/2|i\ell)^2}{\vartheta_1(1/2|i\ell)^2}\tilde{\mathcal{L}}[1,1](\ell)
+4\prod_{i=1}^3\frac{\vartheta_s(v_i|i\ell)}{\vartheta_1(v_i|i\ell)}
-4\prod_{i=1}^3\frac{\vartheta_s(3v_i|i\ell)}{\vartheta_1(3v_i|i\ell)}
\Bigg\}\\
\mathcal{Z}_{\mathcal{K}}&= 16\frac{V_4}{(16\pi^2)^2}
\sum_{s=1}^4(-)^{s-1}\int_0^{\infty}d\ell\frac{\vartheta_s(0|i\ell)}{\eta(i\ell)^{3}}
\Bigg\{\frac{\vartheta_s(0|i\ell)^3}{\eta(i\ell)^{9}}\mathcal{L}_{\mathcal{K}}(\ell)\nonumber\\
&\qquad\qquad - \frac{\vartheta_s(0|il)}{\eta(i\ell)^3}
\frac{\vartheta_s(1/2|i\ell)^2}{\vartheta_1(1/2|i\ell)^2}\tilde{\mathcal{L}}[2,2](\ell)
+4\prod_{i=1}^3\frac{\vartheta_s(v_i|i\ell)}{\vartheta_1(v_i|i\ell)}
-4\prod_{i=1}^3\frac{\vartheta_s(3v_i|i\ell)}{\vartheta_1(3v_i|i\ell)}
\Bigg\}\\
\mathcal{Z}_{\mathcal{M}}&= -2M\frac{V_4}{(16\pi^2)^2}
\sum_{s=1}^4(-)^{s-1}\int_0^{\infty}d\ell\frac{\vartheta_s(0|-1/2+i\ell)}{\eta(-1/2+i\ell)^{3}}
\Bigg\{\frac{\vartheta_s(0|-1/2+i\ell)^3}{\eta(-1/2+i\ell)^{9}}\mathcal{L}_{\mathcal{M}}(\ell)\nonumber\\
&\qquad\qquad\qquad\qquad
-\frac{\vartheta_s(0|-1/2+i\ell)}{\eta(-1/2+i\ell)^3}
\frac{\vartheta_s(1/2|-1/2+i\ell)^2}{\vartheta_1(1/2|-1/2+i\ell)^2}\tilde{\mathcal{L}}[2,2](\ell)\nonumber\\
&\qquad\qquad\qquad\qquad
+4\prod_{i=1}^3\frac{\vartheta_s(v_i|-1/2+i\ell)}{\vartheta_1(v_i|-1/2+i\ell)}
-4\prod_{i=1}^3\frac{\vartheta_s(3v_i|-1/2+i\ell)}{\vartheta_1(3v_i|-1/2+i\ell)}
\Bigg\}.
\end{align}

Again, the precise form of the lattice factors
$\mathcal{L}_{\sigma}$ coming from untwisted fields in the loop
channel will be irrelevant. On the other hand, the lattice factors
coming from $\mathbb{Z}_2$ twisted strings in the loop channel are
relevant. Their explicit form is given by

\begin{equation}
\tilde{\mathcal{L}}[\alpha,\beta](\ell)=\left(\sum_{m\in\mathbb{Z}}e^{-\alpha\pi
\ell m^2R^2/2}\right)\left(\sum_{n\in\mathbb{Z}}e^{-2\beta\pi \ell
n^2/R^2}\right),
\end{equation}

where $R$ is the compactification radius (for simplicity, all six
radii are taken to be equal). Tadpole cancellation requires
$M=16$. Then, using equations \eqref{thetaatzero}, \eqref{Jacobi1}
and \eqref{Jacobi2}, we find

\begin{align}
\delta_{\mathcal{A}}&=+\frac{g_s^2}{8\pi^2}\left[\int_0^{\infty}d\ell\ell\sum_{i=1}^3\frac{\vartheta_1'(v_i|i\ell)}{\vartheta_1(v_i|i\ell)}
+\frac{\pi}{4}\int_0^{\infty}d\ell\ell\mathcal{L}[1,1](\ell)\right]\\
\delta_{\mathcal{K}}&=+\frac{g_s^2}{8\pi^2}\left[\int_0^{\infty}d\ell\ell\sum_{i=1}^3\frac{\vartheta_1'(v_i|i\ell)}{\vartheta_1(v_i|i\ell)}
+\frac{\pi}{4}\int_0^{\infty}d\ell\ell\mathcal{L}[2,2](\ell)\right]\\
\delta_{\mathcal{M}}&=-\frac{g_s^2}{4\pi^2}\left[\int_0^{\infty}d\ell\ell\sum_{i=1}^3
\frac{\vartheta_1'(v_i|-1/2+i\ell)}{\vartheta_1(v_i|-1/2+i\ell)}
+\frac{\pi}{4}\int_0^{\infty}d\ell\ell\mathcal{L}[2,2](\ell)\right],
\end{align}

and therefore, the induced Planck mass becomes

\begin{align}
\delta=&\frac{g_s^2}{4\pi^2}\int_0^{\infty}d\ell\ell\sum_{i=1}^3\left(
\frac{\vartheta_1'(v_i|i\ell)}{\vartheta_1(v_i|i\ell)}-\frac{\vartheta_1'(v_i|-1/2+i\ell)}{\vartheta_1(v_i|-1/2+i\ell)}\right)\nonumber\\
&+\frac{g_s^2}{32\pi}\int_0^{\infty}d\ell\ell\left(\mathcal{L}[1,1](\ell)-\mathcal{L}[2,2](\ell)\right).
\end{align}

The first term comes from the untwisted sector and is of the same
form as the corresponding term for the $\mathbb{Z}_3$-model.
Evaluating the integral gives a contribution of the form (again,
reinserting a factor of $2/\alphaprime$):

\begin{equation}
\delta_{\text{untwisted}}=\frac{1}{2^8\pi^2}\left((\zeta(2,1/4)-\zeta(2,3/4)\right)g_s^2M_s^2\approx
5.8\times 10^{-3}g_s^2M_s^2.
\end{equation}

On the other hand, the second term is derived from the
$\mathbb{Z}_2$-twisted sector. In this case, evaluating the
integral gives the following expression (see appendix
\ref{usefulintegrals}):

\begin{equation}
\delta_{\text{twisted}}=
\frac{3}{32\pi^3}\left(\sum_{n,m=1}^{\infty}\frac{1}{\left(\frac{1}{2}(mR)^2+2\left(\frac{n}{R}\right)^2\right)^2}
+\frac{\pi^4}{180}\left(\frac{R^2}{2}+\frac{2}{R^2}\right)\right)M_s^2.
\end{equation}

Because the $\mathbb{Z}_2$-subgroup of the orbifold group leaves
the third torus invariant, the $R\rightarrow\frac{\alphaprime}{R}$
T-duality reappears in the $\mathbb{Z}_2$-twisted sector. For $R$
much bigger than one, the term which is proportional to $R^2$
dominates over all other contributions to the induced Planck mass.
We can write:

\begin{equation}
\delta=\frac{\pi}{2^6\cdot 3\cdot
5}R^2g_s^2M_s^4+\mathcal{O}(R^0).
\end{equation}

We recognize a Kaluza-Klein like dependence on the
compactification scale. In fact, the $\mathbb{Z}_2$-twisted sector
consists of open and closed string states which are localized on a
six-dimensional surface. We can interpret the above contribution
to the four-dimensional Planck mass as coming from an induced
six-dimensional Planck mass which is subsequently reduced to four
dimensions by a Kaluza-Klein mechanism.

\section{Conclusions and discussion}
\label{conclusions}

Quantum loop corrections generically renormalize the
four-dimensional Planck mass in brane world models. We have shown
how these corrections arise in supersymmetric orientifold models
with intersecting branes. For $\Omega\mathcal{R}$-orientifolds
with local tadpole cancellation, we found that tadpole
cancellation implies a UV-finite graviton scattering amplitude and
we have explicitly computed the induced four-dimensional Planck
mass in this case. We would like to stress that, in principle,
computing the induced Planck mass is completely under control for
all supersymmetric Intersecting Brane Worlds. One can use the
formuli \eqref{treedelta1} to \eqref{treedelta3} as a black box
taking only the open string partition function as an input.
However, UV-divergencies in graviton scattering amplitudes could
potentially spoil perturbative calculability. It remains an open
question as to whether this happens in the context of graviton
scattering in the presence of D-branes and O-planes.

DGP-gravity allows us in principle to think about non-compact
extra dimensions. However, one of the most attractive features of
intersecting brane worlds is that for compactified extra
dimensions, two branes can multiply intersect with each other,
giving rise to family replication. Also, branes wrapping infinite
volume cycles in the extra dimensions don't support local gauge
symmetries as their four-dimensional coupling constant is
proportional to the branes' volume in the extra dimensions. If we
stick to compact extra dimensions, the four-dimensional Planck
mass gets a Kaluza-Klein contribution from the ten-dimensional
Planck mass which is of the form

\begin{equation}
M_{P,KK}^2 \sim M_s^8 V_6,
\end{equation}

where $V_6$ is the compactification volume. For supersymmetric
string vacua, it is natural to take the string scale to be of the
order of the GUT scale and as a consequence, the Kaluza-Klein
mechanism accounts for the observed Planck mass if the extra
dimensions are moderately enlarged. Induced gravity doesn't change
the picture significantly. However, there are at least two reasons
why DGP-gravity might be important in intersecting brane worlds.
First, a bottom-up approach to intersecting brane worlds has been
proposed \cite{Uranga:2002pg} in which branes wrap small cycles in
a non-compact Calabi-Yau 3-fold. There, an induced
Einstein-Hilbert term on the brane intersections would localize
gravity and explain the observed $1/r^2$-dependence of the
gravitational force. Second, and perhaps more importantly, one of
the most exciting prospects of future table-top experiments is the
possibility of finding large extra dimensions. In
non-supersymmetric models without induced gravity, these could
well be in the sub-millimeter regime. Constraints on the size of
the extra dimensions are derived from the energy loss that high
energy events on the brane, such as supernovae or collider
experiments, would suffer due to emission of light Kaluza-Klein
modes of the graviton into the bulk. Because in DGP-gravity,
graviton emission is strongly suppressed for energies greater than
a cross-over scale, which is controlled by the magnitude of the
induced Planck mass and the effective width of the brane, the
constraints on the size of extra dimensions can be relaxed
considerably \cite{Dvali:2001gm, Kiritsis:2003mc}. While this
argument is not relevant for toroidal orientifolds with
intersecting $D6$-branes\footnote{For intersecting D6-branes
wrapping $T^6$, the maximum size of the extra dimensions is more
seriously constrained by the non-observation of Kaluza-Klein modes
of the Standard model gauge fields as there is no direction which
is transverse to all branes}, it applies to more general
backgrounds with large transverse directions and to orientifolds
with intersecting $D4$-branes or $D5$-branes.

The specific models we examined in this work are
phenomenologically unappealing, being non-chiral. However, we
regard our results on induced gravity as simple examples of a more
general mechanism. In this respect it is interesting that it is
possible to get an induced Planck mass which grows linearly with
the volume of some extra dimensions. This means that for large
extra dimensions, the pure Kaluza-Klein contribution to the
four-dimensional Planck mass does not necessarily dominate over
the induced Einstein-Hilbert term.

As we pointed out, it is not clear how tadpole cancellation
insures UV-finiteness of the graviton scattering amplitude in more
general backgrounds. In general, the divergencies in the one-loop
graviton amplitude have a very different structure from those in
the vacuum amplitude. For example, two D6-branes which intersect
at angles $\phi_i, i=1\dots 3$, lead to a term of the form:

\begin{equation}
\frac{\vartheta_s(0)}{\eta^3}\prod_{i=1}^{3}\frac{\vartheta_s(\phi_i/\pi)}{\vartheta_1(\phi_i/\pi)}
\end{equation}

in the tree level partition function. As can easily be checked,
this term contributes to the RR-tadpole divergence in the vacuum
amplitude in the following form:

\begin{equation}
\sim \prod_{i=1}^3 \cot(\phi_i)\int_0^{\infty}d\ell.
\end{equation}

On the other hand, it contributes to the divergence in the
graviton scattering amplitude through a term

\begin{equation}
\sim \sum_{i=1}^3 \cot(\phi_i)\int_0^{\infty}d\ell\ell.
\end{equation}

If we consider orientifold models where D-branes are not aligned
to the O-planes, the partition functions in the $\mathcal{A}$-,
$\mathcal{K}$- and $\mathcal{M}$-sector don't have a mutually
identical structure any longer. Instead, they conspire
non-trivially to cancel the infinities in the one-loop vacuum
amplitude. It is therefore not obvious whether UV-finiteness holds
for the graviton scattering amplitude. In the absence of any
general theorems about finiteness of open string scattering
amplitudes, it would be interesting to have some empirical
evidence instead. For a few simple examples, we have checked that
the graviton scattering amplitude remains finite by an explicit
computation. However, we are unable at the moment to give a
general argument in favour of finiteness. We hope to report on
this issue in the future.

Finally, we would like to point out that an interesting extension
of our work would be to compute moduli scattering in toroidal
$\mathcal{N}=1$ string compactifications and thus determine
one-loop corrections to the K\"{a}hler potential. A deviation from
the no-scale form would be highly interesting in the context of
moduli stabilization.

\acknowledgments

It is a pleasure to thank Fernando Quevedo, Paul Davis and Ralph
Blumenhagen for their help in various stages of this project. I
also want to thank Stephen Hawking for his patience and
commitment. This work was supported by EPSRC and a Gates Cambridge
Scholarship.

\appendix

\section{Jacobi Theta Functions}
\label{Jacobi}

The Jacobi Theta functions are defined by

\begin{align}
&\vartheta_1(z|\tau)=\sum_{n\in\mathbb{Z}}q^{(n+1/2)^2/2}e^{2\pi i
(n+1/2)(z+1/2)},&&
\vartheta_2(z|\tau)=\sum_{n\in\mathbb{Z}}q^{(n+1/2)^2/2}e^{2\pi i
(n+1/2)z}\\
&\vartheta_3(z|\tau)=\sum_{n\in\mathbb{Z}}q^{n^2/2}e^{2\pi i
nz},&& \vartheta_4(z|\tau)=\sum_{n\in\mathbb{Z}}q^{n^2/2}e^{2\pi i
nz+1/2)}
\end{align}

where $q=e^{2\pi i\tau}$ is called the nome and the multi-valued
function $q^\lambda$ is interpreted to stand for $e^{2\pi i
\tau\lambda}$. They have alternative product representations

\begin{align}
\vartheta_1(z|\tau)&=i\eta(\tau)e^{\pi iz}
q^{\frac{1}{12}}\prod_{n=1}^{\infty}
\left(1-e^{2\pi iz}q^{n}\right) \left(1-e^{-2\pi iz}q^{n-1}\right)\\
\vartheta_2(z|\tau)&=\eta(\tau)e^{\pi iz}
q^{\frac{1}{12}}\prod_{n=1}^{\infty}
\left(1+e^{2\pi iz}q^{n}\right) \left(1+e^{-2\pi iz}q^{n-1}\right)\\
\vartheta_3(z|\tau)&=\eta(\tau)q^{-\frac{1}{24}}\prod_{n=1}^{\infty}
\left(1+e^{2\pi iz}q^{n-\frac{1}{2}}\right)\left(1+e^{-2\pi iz}q^{n-\frac{1}{2}}\right)\\
\vartheta_4(z|\tau)&=\eta(\tau)q^{-\frac{1}{24}}\prod_{n=1}^{\infty}
\left(1-e^{2\pi iz}q^{n-\frac{1}{2}}\right)\left(1-e^{-2\pi
iz}q^{n-\frac{1}{2}}\right),
\end{align}

where $\eta$ is the Dedekind function

\begin{equation}
\eta(\tau)=q^{1/24}\prod_{n=1}^{\infty}(1-q^n).
\end{equation}

By convention, the derivative is taken with respect to the first
variable. The following modular transformations hold:

\begin{align}
\label{modular1}
\vartheta_1(z|\tau)&=(-i\tau)^{-1/2}\exp\left({-\frac{i\pi
z^2}{\tau}}\right)i\vartheta_1\left(\frac{z}{\tau}\Big|-\frac{1}{\tau}\right)\\
\vartheta_2(z|\tau)&=(-i\tau)^{-1/2}\exp\left({-\frac{i\pi
z^2}{\tau}}\right)\vartheta_4\left(\frac{z}{\tau}\Big|-\frac{1}{\tau}\right)\\
\vartheta_3(z|\tau)&=(-i\tau)^{-1/2}\exp\left({-\frac{i\pi
z^2}{\tau}}\right)\vartheta_3\left(\frac{z}{\tau}\Big|-\frac{1}{\tau}\right)\\
\vartheta_4(z|\tau)&=(-i\tau)^{-1/2}\exp\left({-\frac{i\pi
z^2}{\tau}}\right)\vartheta_2\left(\frac{z}{\tau}\Big|-\frac{1}{\tau}\right)
\end{align}

for S-transformations and

\begin{align}
&\vartheta_1(z|\tau)=e^{i\pi/4}\vartheta_1(z|\tau+1); && \vartheta_2(z|\tau)=e^{i\pi/4}\vartheta_2(z|\tau+1)\\
\label{modular2} &\vartheta_3(z|\tau)=\vartheta_4(z|\tau+1); &&
\vartheta_4(z|\tau)=\vartheta_3(z|\tau+1)\\
\intertext{for T-transformations. The Dedekind function obeys}
&\eta(\tau)=(-i\tau)^{-1/2}\eta(-1/\tau); &&
\eta(\tau)=e^{-i\pi/12}\eta(\tau+1).
\end{align}

The ratios of derivatives of Jacobi theta functions to the
functions themselves have a simple form \cite{Whittaker}. In
particular, we have:

\begin{equation}
\label{JacobiPrime}
\frac{\vartheta_1'(z|\tau)}{\vartheta_1(z|\tau)}=\cot(\pi
z)+4\sum_{n=1}^{\infty}\frac{q^n}{1-q^n} \sin(2\pi nz).
\end{equation}

The first theta function also satisfies the following equations:

\begin{equation}
\label{thetaatzero}
\vartheta_1(0|\tau)=0,\qquad
\vartheta_1'(0|\tau)=-2\pi\eta(\tau)^3,\qquad
\vartheta_1''(0|\tau)=0.
\end{equation}

The Jacobi theta functions are solutions to the heat equation

\begin{equation}
\partial_z^2\vartheta_s(z|\tau)=4\pi
i\partial_{\tau}\vartheta_s(z|\tau).
\end{equation}

They satisfy a great number of identities which can be viewed as
generalizations of Jacobi's `abstruse identity'. Most useful to us
is the following:

\begin{equation}
\sum_{s=1}^4(-)^{s-1}\prod_{i=1}^4\vartheta_s(v_i)=2\prod_{i=1}^4\vartheta_1(v_i')
\end{equation}

where one defines

\begin{align}
& v_1'=\frac{1}{2}(v_1+v_2+v_3+v_4); &&
v_2'=\frac{1}{2}(v_1+v_2-v_3-v_4)\\
& v_3'=\frac{1}{2}(v_1-v_2+v_3-v_4); &&
v_4'=\frac{1}{2}(v_1-v_2-v_3+v_4).
\end{align}

and we omit the second argument of the theta functions. For the
special case of $v_1+v_2+v_3=0$, we get the following relation:

\begin{equation}
\label{JacobiIdentity}
\sum_{s=2,3,4}(-)^{s-1}\left\{\vartheta_s(v)\prod_{i=1}^3\vartheta_s(v_i)\right\}=
-\vartheta_1(v)\prod_{i=1}^3\vartheta_1(v_i)+2\vartheta_1\left(\frac{v}{2}\right)\prod_{i=1}^3
\vartheta_1\left(\frac{v}{2}+v_i\right).
\end{equation}

In particular, using equations \eqref{thetaatzero} and the
antisymmetry of $\theta_1(z|\tau)$ about $z=0$, we find:

\begin{equation}
\label{Jacobi1}
\partial_v^2\sum_{s=2,3,4}(-)^{s-1}\left\{\vartheta_s(v)\prod_{i=1}^3\vartheta_s(v_i)\right\}\Bigg|_{v=0}=
-\pi\eta(\tau)^3\sum_{i=1}^3\frac{\vartheta_1'(v_i)}{\vartheta_1(v_i)}\prod_{j=1}^3\vartheta_1(v_j).
\end{equation}

This expression is well-defined if none of the $v_i$ vanish
individually. On the other hand, we find that for $v_1=v_2+v_3=0$,
the following equation holds:

\begin{equation}
\label{Jacobi2}
\partial_v^2\sum_{s=2,3,4}(-)^{s-1}\left\{\vartheta_s(v)\prod_{i=1}^3\vartheta_s(v_i)\right\}\Bigg|_{v=0}=
2\pi^2\eta(\tau)^6\prod_{j=2,3}\vartheta_1(v_j).
\end{equation}

\section{Some useful integrals}
\label{usefulintegrals}

In this appendix, we evaluate two integrals which appear in the
computation of the graviton scattering amplitude. The first
integral is given by

\begin{equation}
I_1(z)=\int_0^{\infty} d\ell\ \ell
\left(\frac{\vartheta_1'(z|\tau)}{\vartheta_1(z|\tau)}-\frac{\vartheta_1'(z|-1/2+\tau)}{\vartheta_1(z|-1/2+\tau)}\right).
\end{equation}

If we define $q\equiv\exp(2\pi\tau)$, we can write (using
\eqref{JacobiPrime}):

\begin{align}
I_1&=\int_0^{\infty} d\ell\ \ell \sum_{n=1}^{\infty}\sin(2\pi
nz)\left(\frac{q^n}{1-q^n}-\frac{(-q)^n}{1-(-q)^n}\right)\\
&=\frac{1}{2\pi^2}\sum_{n\text{ odd}}\frac{\sin(2\pi
nz)}{n^2}\int_0^{\infty}
d\ell \ \ell\frac{e^{-\ell}}{1-e^{-2\ell}}\\
&=\frac{1}{16}\sum_{n\text{ odd}}\frac{\sin(2\pi nz)}{n^2}.
\end{align}

This expression can be evaluated in terms of a finite sum of
Hurwitz zeta functions,

\begin{equation}
\zeta(s,a)=\sum_{n=0}^{\infty}\frac{1}{(n+a)^s},
\end{equation}

because the integral in question always appears with rational
argument. For instance, in the $\mathbb{Z}_3$- and
$\mathbb{Z}_6$-models we have to compute

\begin{align}
I_1(1/3)&=\frac{1}{16}\sum_{n\text{ odd}}\frac{\sin(2\pi
n/3)}{n^2}\\
&=\frac{1}{16}\sin(2\pi/3)\sum_{n=0}^{\infty}\left(\frac{1}{(6n+1)^2}-\frac{1}{(6n+5)^2}\right)\\
&=\frac{\sqrt{3}}{32\cdot 36}(\zeta(2,1/6)-\zeta(2,5/6)).
\end{align}

The second integral we are interested in appears whenever the
orbifold group has a $\mathbb{Z}_2$ subgroup. It is given by

\begin{equation}
I_2(a,R)=\int_0^{\infty} d\ell\
\ell\sum_{m,n\in\mathbb{Z}}e^{-a\ell\pi(m^2R^2/2+2n^2/R^2)}.
\end{equation}

This integral is divergent due to the $m=n=0$ part of the sum.
However, we can formally separate the divergence which will cancel
in the full scattering amplitude and focus on the convergent part
of the integral:

\begin{align}
\tilde{I}_2(a,R)&=\int_0^{\infty} d\ell \
\ell\sum_{\substack{n,m\in\mathbb{Z}\\(n,m)\neq(0,0)}}e^{-a\ell\pi(m^2R^2/2+2n^2/R^2)}\\
&=\int_0^{\infty} d\ell \ \ell\left(4\sum_{n,m=1}^{\infty}
e^{-a\ell\pi(m^2R^2/2+2n^2/R^2)} +2\sum_{m=1}^{\infty}e^{-a\ell\pi
m^2R^2/2}
+2\sum_{n=1}^{\infty}e^{-a\ell\pi 2n^2/R^2}\right)\\
&=\frac{4}{a^2\pi^2}\sum_{n,m=1}^{\infty}\frac{1}{(m^2R^2/2+2(n/R)^2)^2}
+\frac{2\zeta(4)}{a^2\pi^2}\left(\frac{R^2}{2}+\frac{2}{R^2}\right).
\end{align}

It is interesting to consider the case where $R\gg 1$. In this
limit, we can approximate the sum over $n$ by an integral, making
the following substitutions:

\begin{equation}
\frac{1}{R}=\Delta x,\qquad \frac{n}{R}= x,\qquad
\sum_{n=1}^{\infty}\approx R\int_0^{\infty} dx.
\end{equation}

From this, it follows that

\begin{equation}
\sum_{n,m=1}^{\infty}\frac{1}{(m^2R^2/2+2(n/R)^2)^2}\approx
R\sum_{n,m=1}^{\infty}\int_0^{\infty}\frac{dx}{(m^2R^2/2+2x^2)^2}=
\frac{\pi}{2R^2}\zeta(3).
\end{equation}

Therefore, we can write:

\begin{equation}
\tilde{I}(a,R)=\frac{\zeta(4)}{a^2\pi^2}R^2+\mathcal{O}(1/R^2).
\end{equation}


\bibliographystyle{JHEP-2}

\begin{thebibliography}{10}

\bibitem{IBW}
R.~Blumenhagen, L.~Goerlich, B.~Kors and D.~Lust, {\it
Noncommutative
  compactifications of type i strings on tori with magnetic background flux},
  {\em JHEP} {\bf 10} (2000) 006
  [\href{http://arXiv.org/abs/hep-th/0007024}{{\tt hep-th/0007024}}].

G.~Aldazabal, S.~Franco, L.~E. Ibanez, R.~Rabadan and A.~M.
Uranga, {\it
  Intersecting brane worlds},  {\em JHEP} {\bf 02} (2001) 047
  [\href{http://arXiv.org/abs/hep-ph/0011132}{{\tt hep-ph/0011132}}].

G.~Aldazabal, S.~Franco, L.~E. Ibanez, R.~Rabadan and A.~M.
Uranga, {\it D = 4
  chiral string compactifications from intersecting branes},  {\em J. Math.
  Phys.} {\bf 42} (2001) 3103--3126
  [\href{http://arXiv.org/abs/hep-th/0011073}{{\tt hep-th/0011073}}].

R.~Blumenhagen, B.~Kors and D.~Lust, {\it Type i strings with f-
and b-flux},
  {\em JHEP} {\bf 02} (2001) 030
  [\href{http://arXiv.org/abs/hep-th/0012156}{{\tt hep-th/0012156}}].

L.~E. Ibanez, F.~Marchesano and R.~Rabadan, {\it Getting just the
standard
  model at intersecting branes},  {\em JHEP} {\bf 11} (2001) 002
  [\href{http://arXiv.org/abs/hep-th/0105155}{{\tt hep-th/0105155}}].

R.~Blumenhagen, B.~Kors, D.~Lust and T.~Ott, {\it The standard
model from
  stable intersecting brane world orbifolds},  {\em Nucl. Phys.} {\bf B616}
  (2001) 3--33 [\href{http://arXiv.org/abs/hep-th/0107138}{{\tt
  hep-th/0107138}}].

C.~Kokorelis, {\it GUT model hierarchies from intersecting
branes}, {\em JHEP} {\bf 08} (2002) 018
[\href{http://arXiv.org/abs/hep-th/0203187}{{\tt
hep-th/0203187}}].

\bibitem{SusyOfolds}
M.~Cvetic, G.~Shiu and A.~M. Uranga, {\it Chiral four-dimensional
n = 1
  supersymmetric type iia orientifolds from intersecting d6-branes},  {\em
  Nucl. Phys.} {\bf B615} (2001) 3--32
  [\href{http://arXiv.org/abs/hep-th/0107166}{{\tt hep-th/0107166}}].

R.~Blumenhagen, V.~Braun, B.~Kors and D.~Lust, {\it Orientifolds
of k3 and
  calabi-yau manifolds with intersecting d-branes},  {\em JHEP} {\bf 07} (2002)
  026 [\href{http://arXiv.org/abs/hep-th/0206038}{{\tt hep-th/0206038}}].

M.~Cvetic and I.~Papadimitriou, {\it More supersymmetric
standard-like models
  from intersecting d6-branes on type iia orientifolds},  {\em Phys. Rev.} {\bf
  D67} (2003) 126006 [\href{http://arXiv.org/abs/hep-th/0303197}{{\tt
  hep-th/0303197}}].

S.~Forste, G.~Honecker and R.~Schreyer, {\it Supersymmetric z(n) x
z(m)
  orientifolds in 4d with d-branes at angles},  {\em Nucl. Phys.} {\bf B593}
  (2001) 127--154 [\href{http://arXiv.org/abs/hep-th/0008250}{{\tt
  hep-th/0008250}}].

G.~Honecker, {\it Chiral supersymmetric models on an orientifold
of z(4) x z(2)
  with intersecting d6-branes},  {\em Nucl. Phys.} {\bf B666} (2003) 175--196
  [\href{http://arXiv.org/abs/hep-th/0303015}{{\tt hep-th/0303015}}].

G.~Honecker and T.~Ott, {\it Getting just the supersymmetric
standard model at
  intersecting branes on the z(6)-orientifold},
  \href{http://arXiv.org/abs/hep-th/0404055}{{\tt hep-th/0404055}}.

G.~Honecker, {\it Chiral n = 1 4d orientifolds with d-branes at
angles},  {\em
  Mod. Phys. Lett.} {\bf A19} (2004) 1863--1879
  [\href{http://arXiv.org/abs/hep-th/0407181}{{\tt hep-th/0407181}}].

C.~Kokorelis, {\it N = 1 supersymmetric standard models from
intersecting branes},
  \href{http://arXiv.org/abs/hep-th/0406258}{{\tt hep-th/0406258}}.

M.~Cvetic, T.~Li and T.~Liu, {\it supersymmetric pati-salam models
from intersecting d6-branes: a road to the standard model},
  \href{http://arXiv.org/abs/hep-th/0403061}{{\tt hep-th/0403061}}.

M.~Cvetic, P.~Langacker, T.~J.~Li and T.~Liu, {\it d6-brane
splitting on type IIA orientifolds},
  \href{http://arXiv.org/abs/hep-th/0407178}{{\tt hep-th/0407178}}.



\bibitem{Lust:2004ks}
D.~Lust, {\it Intersecting brane worlds: A path to the standard
model?},
  \href{http://arXiv.org/abs/hep-th/0401156}{{\tt hep-th/0401156}}.

\bibitem{Kiritsis:2003mc}
E.~Kiritsis, {\it D-branes in standard model building, gravity and
cosmology},
  {\em Fortsch. Phys.} {\bf 52} (2004) 200--263
  [\href{http://arXiv.org/abs/hep-th/0310001}{{\tt hep-th/0310001}}].

\bibitem{Lust:2003ky}
D.~Lust and S.~Stieberger, {\it Gauge threshold corrections in
intersecting
  brane world models},  \href{http://arXiv.org/abs/hep-th/0302221}{{\tt
  hep-th/0302221}}.

\bibitem{Blumenhagen:2003jy}
R.~Blumenhagen, D.~Lust and S.~Stieberger, {\it Gauge unification
in
  supersymmetric intersecting brane worlds},  {\em JHEP} {\bf 07} (2003) 036
  [\href{http://arXiv.org/abs/hep-th/0305146}{{\tt hep-th/0305146}}].

\bibitem{Cvetic:2003ch}
M.~Cvetic and I.~Papadimitriou, {\it Conformal field theory
couplings for
  intersecting d-branes on orientifolds},  {\em Phys. Rev.} {\bf D68} (2003)
  046001 [\href{http://arXiv.org/abs/hep-th/0303083}{{\tt hep-th/0303083}}].

\bibitem{Cremades:2003qj}
D.~Cremades, L.~E. Ibanez and F.~Marchesano, {\it Yukawa couplings
in
  intersecting d-brane models},  {\em JHEP} {\bf 07} (2003) 038
  [\href{http://arXiv.org/abs/hep-th/0302105}{{\tt hep-th/0302105}}].

\bibitem{Abel:2003vv}
S.~A. Abel and A.~W. Owen, {\it Interactions in intersecting brane
models},
  {\em Nucl. Phys.} {\bf B663} (2003) 197--214
  [\href{http://arXiv.org/abs/hep-th/0303124}{{\tt hep-th/0303124}}].

\bibitem{Abel:2003yx}
S.~A. Abel and A.~W. Owen, {\it N-point amplitudes in intersecting
brane
  models},  {\em Nucl. Phys.} {\bf B682} (2004) 183--216
  [\href{http://arXiv.org/abs/hep-th/0310257}{{\tt hep-th/0310257}}].

\bibitem{Hashimoto:2003xz}
K.~Hashimoto and S.~Nagaoka, {\it Recombination of intersecting
d-branes by
  local tachyon condensation},  {\em JHEP} {\bf 06} (2003) 034
  [\href{http://arXiv.org/abs/hep-th/0303204}{{\tt hep-th/0303204}}].

\bibitem{Jones:2003ew}
N.~T. Jones and S.~H.~H. Tye, {\it Spectral flow and boundary
string field
  theory for angled d-branes},  {\em JHEP} {\bf 08} (2003) 037
  [\href{http://arXiv.org/abs/hep-th/0307092}{{\tt hep-th/0307092}}].

\bibitem{Epple:2003xt}
F.~Epple and D.~Lust, {\it Tachyon condensation for intersecting
branes at
  small and large angles},  {\em Fortsch. Phys.} {\bf 52} (2004) 367--387
  [\href{http://arXiv.org/abs/hep-th/0311182}{{\tt hep-th/0311182}}].

\bibitem{Nagaoka:2003zn}
S.~Nagaoka, {\it Higher dimensional recombination of intersecting
d-branes},
  {\em JHEP} {\bf 02} (2004) 063
  [\href{http://arXiv.org/abs/hep-th/0312010}{{\tt hep-th/0312010}}].

\bibitem{Lust:2004cx}
D.~Lust, P.~Mayr, R.~Richter and S.~Stieberger, {\it Scattering of
gauge,
  matter, and moduli fields from intersecting branes},
  \href{http://arXiv.org/abs/hep-th/0404134}{{\tt hep-th/0404134}}.

\bibitem{Dvali:2000hr}
G.~R. Dvali, G.~Gabadadze and M.~Porrati, {\it 4d gravity on a
brane in 5d
  minkowski space},  {\em Phys. Lett.} {\bf B485} (2000) 208--214
  [\href{http://arXiv.org/abs/hep-th/0005016}{{\tt hep-th/0005016}}].

\bibitem{Dvali:2000xg}
G.~R. Dvali and G.~Gabadadze, {\it Gravity on a brane in
infinite-volume extra
  space},  {\em Phys. Rev.} {\bf D63} (2001) 065007
  [\href{http://arXiv.org/abs/hep-th/0008054}{{\tt hep-th/0008054}}].

\bibitem{Adler:1982ri}
S.~L. Adler, {\it Einstein gravity as a symmetry breaking effect
in quantum
  field theory},  {\em Rev. Mod. Phys.} {\bf 54} (1982) 729.

\bibitem{Bachas:1999um}
C.~P. Bachas, P.~Bain and M.~B. Green, {\it Curvature terms in
d-brane actions
  and their m-theory origin},  {\em JHEP} {\bf 05} (1999) 011
  [\href{http://arXiv.org/abs/hep-th/9903210}{{\tt hep-th/9903210}}].

\bibitem{Schnitzer:2002rt}
H.~J. Schnitzer and N.~Wyllard, {\it An orientifold of ads(5) x
t(11) with
  d7-branes, the associated alpha'**2 corrections and their role in the dual n
  = 1 sp(2n+2m) x sp(2n) gauge theory},  {\em JHEP} {\bf 08} (2002) 012
  [\href{http://arXiv.org/abs/hep-th/0206071}{{\tt hep-th/0206071}}].

\bibitem{Kiritsis:1995ta}
E.~Kiritsis and C.~Kounnas, {\it Infrared regularization of
superstring theory
  and the one loop calculation of coupling constants},  {\em Nucl. Phys.} {\bf
  B442} (1995) 472--493 [\href{http://arXiv.org/abs/hep-th/9501020}{{\tt
  hep-th/9501020}}].

\bibitem{Kiritsis:2001bc}
E.~Kiritsis, N.~Tetradis and T.~N.~Tomaras, {\it Thick branes and
4D gravity}, {\em JHEP} {\bf 08} (2001) 012
[\href{http://arXiv.org/abs/hep-th/0106050}{{\tt
arXiv:hep-th/0106050}}].

\bibitem{Antoniadis:1997eg}
I.~Antoniadis, S.~Ferrara, R.~Minasian and K.~S. Narain, {\it R**4
couplings in
  m- and type ii theories on calabi-yau spaces},  {\em Nucl. Phys.} {\bf B507}
  (1997) 571--588 [\href{http://arXiv.org/abs/hep-th/9707013}{{\tt
  hep-th/9707013}}].

\bibitem{Antoniadis:2002tr}
I.~Antoniadis, R.~Minasian and P.~Vanhove, {\it Non-compact
calabi-yau
  manifolds and localized gravity},  {\em Nucl. Phys.} {\bf B648} (2003) 69--93
  [\href{http://arXiv.org/abs/hep-th/0209030}{{\tt hep-th/0209030}}].

\bibitem{Antoniadis:2003sw}
I.~Antoniadis, R.~Minasian, S.~Theisen and P.~Vanhove, {\it String
loop
  corrections to the universal hypermultiplet},  {\em Class. Quant. Grav.} {\bf
  20} (2003) 5079--5102 [\href{http://arXiv.org/abs/hep-th/0307268}{{\tt
  hep-th/0307268}}].

\bibitem{Kakushadze:1998tr}
Z.~Kakushadze, {\it Gauge theories from orientifolds and large n
limit},  {\em
  Nucl. Phys.} {\bf B529} (1998) 157--179
  [\href{http://arXiv.org/abs/hep-th/9803214}{{\tt hep-th/9803214}}].

\bibitem{Kakushadze:2001bd}
Z.~Kakushadze, {\it Orientiworld},  {\em JHEP} {\bf 10} (2001) 031
  [\href{http://arXiv.org/abs/hep-th/0109054}{{\tt hep-th/0109054}}].

\bibitem{Kohlprath:2003pu}
E.~Kohlprath, {\it Induced gravity in z(n) orientifold models},
  \href{http://arXiv.org/abs/hep-th/0311251}{{\tt hep-th/0311251}}.

\bibitem{Polchinski:1996na}
J.~Polchinski, {\it Lectures on d-branes},
  \href{http://arXiv.org/abs/hep-th/9611050}{{\tt hep-th/9611050}}.

\bibitem{Blumenhagen:1999md}
R.~Blumenhagen, L.~Gorlich and B.~Kors, {\it Supersymmetric
orientifolds in 6d
  with d-branes at angles},  {\em Nucl. Phys.} {\bf B569} (2000) 209--228
  [\href{http://arXiv.org/abs/hep-th/9908130}{{\tt hep-th/9908130}}].

\bibitem{Blumenhagen:1999ev}
R.~Blumenhagen, L.~Gorlich and B.~Kors, {\it Supersymmetric 4d
orientifolds of
  type iia with d6-branes at angles},  {\em JHEP} {\bf 01} (2000) 040
  [\href{http://arXiv.org/abs/hep-th/9912204}{{\tt hep-th/9912204}}].

\bibitem{Pradisi:1999ii}
G.~Pradisi, {\it type I vacua from diagonal Z(3)-orbifolds},
Nucl.\ Phys.\ B {\bf 575} (2000) 134,
  \href{http://arXiv.org/abs/hep-th/9912218}{{\tt hep-th/9912218}}.

\bibitem{Blumenhagen:2004di}
R.~Blumenhagen, J.~P. Conlon and K.~Suruliz, {\it Type iia
orientifolds on
  general supersymmetric z(n) orbifolds},
  \href{http://arXiv.org/abs/hep-th/0404254}{{\tt hep-th/0404254}}.

\bibitem{Coleman:1973jx}
S.~R. Coleman and E.~Weinberg, {\it Radiative corrections as the
origin of
  spontaneous symmetry breaking},  {\em Phys. Rev.} {\bf D7} (1973) 1888--1910.

\bibitem{Polchinski:1998rq}
J.~Polchinski, {\it String theory. vol. 1: An introduction to the
bosonic
  string}, . Cambridge, UK: Univ. Pr. (1998) 402 p.

\bibitem{Burgess:1987wt}
C.~P. Burgess and T.~R. Morris, {\it Open superstrings a la
polyakov},  {\em
  Nucl. Phys.} {\bf B291} (1987) 285.

\bibitem{Antoniadis:1997vw}
I.~Antoniadis, C.~Bachas, C.~Fabre, H.~Partouche and T.~R. Taylor,
{\it Aspects
  of type i - type ii - heterotic triality in four dimensions},  {\em Nucl.
  Phys.} {\bf B489} (1997) 160--178
  [\href{http://arXiv.org/abs/hep-th/9608012}{{\tt hep-th/9608012}}].

\bibitem{Angelantonj:1999xf}
C.~Angelantonj and R.~Blumenhagen, {\it Discrete deformations in
type i vacua},
   {\em Phys. Lett.} {\bf B473} (2000) 86--93
  [\href{http://arXiv.org/abs/hep-th/9911190}{{\tt hep-th/9911190}}].

\bibitem{Uranga:2002pg}
A.~M. Uranga, {\it Local models for intersecting brane worlds},
{\em JHEP}
  {\bf 12} (2002) 058 [\href{http://arXiv.org/abs/hep-th/0208014}{{\tt
  hep-th/0208014}}].

\bibitem{Dvali:2001gm}
G.~R. Dvali, G.~Gabadadze, M.~Kolanovic and F.~Nitti, {\it The
power of
  brane-induced gravity},  {\em Phys. Rev.} {\bf D64} (2001) 084004
  [\href{http://arXiv.org/abs/hep-ph/0102216}{{\tt hep-ph/0102216}}].

\bibitem{Whittaker}
E.~T. Whittaker and G.~N. Watson, {\em A Course in Modern
Analysis, 4th ed.}
\newblock Cambridge University Press, 1990.

\end{thebibliography}

\providecommand{\href}[2]{#2}\begingroup\raggedright\endgroup

\end{document}